\documentclass[sigconf]{cidr-2026}

\usepackage{xspace}
\usepackage{amsthm}
\usepackage{multirow}
\usepackage[ruled,vlined,linesnumbered]{algorithm2e}
\usepackage{graphicx}
\usepackage{subcaption}
\usepackage{booktabs}
\usepackage{adjustbox}
\usepackage[dvipsnames]{xcolor}
\usepackage{multicol}

\newcommand{\sys}{\textsc{KathDB}\xspace}

\begin{document}

\title[KathDB: Explainable Multimodal Database Management System with Human-AI Collaboration]{KathDB: Explainable Multimodal Database Management System \\ with Human-AI Collaboration}
\subtitle{Vision Paper}
\renewcommand{\shorttitle}{KathDB: Explainable Multimodal Database Management System with Human-AI Collaboration (Vision Paper)}


\author{Guorui Xiao, Enhao Zhang, Nicole Sullivan, Will Hansen, and Magdalena Balazinska}
\affiliation{%
  \institution{University of Washington}
  \city{Seattle}
  \state{WA}
  \country{USA}
}
\email{{grxiao,enhaoz,nsulliv,willnh,magda}@cs.washington.edu}



\begin{abstract}
Traditional DBMSs execute user- or application-provided SQL queries over relational data with strong semantic guarantees and advanced query optimization, but writing complex SQL is hard and focuses only on structured tables. Contemporary multimodal systems (which operate over relations but also text, images, and even videos) either expose low-level controls that force users to use (and possibly create) machine learning UDFs manually within SQL or offload execution entirely to black-box LLMs, sacrificing usability or explainability. We propose \sys, a new system that combines relational semantics with the reasoning power of foundation models over multimodal data. Furthermore, \sys includes human-AI interaction channels during query parsing, execution, and result explanation, such that users can iteratively obtain explainable answers across data modalities.



\end{abstract}


\maketitle


\section{Introduction}
\label{sec:intro}


Traditional relational database management systems (DBMSs), such as PostgreSQL\footnote{\url{https://www.postgresql.org}} and MySQL\footnote{\url{https://www.mysql.com/}}, were designed to store tabular data and answer queries expressed in SQL.  
Their query optimizers rely on relational algebra and a cost–based model, providing clear query semantics and high efficiency~\cite{chaudhuri1998overview}.
Yet these systems offer no native support for other data modalities (text, images, audio, \ldots), which modern data‑intensive applications in science, healthcare, industry, and media now regard as first‑class citizens~\cite{urban2024caesura,urban2024eleet, kurt2024xmode, liu2025palimpzest}. 
Consider the following natural language (NL) query:
\emph{"Sort the films in the table by how exciting they are, but the poster should be 'boring'."} 
The database consists of a relational table that stores movie metadata such as title and release year, along with a column containing the movie's plot text and another column holding its poster image, represented either by pixel values or, more commonly, by a file path to the image stored on disk.
A classical DBMS cannot evaluate such a query, because computing "excitement" scores over plots requires interpreting unstructured text while labeling posters as "boring" requires understanding poster images. 
Additionally, the NL query is ambiguous about the user's intent: does the user want to filter
posters based on whether they are boring or should
boring be part of the ranking? Such ambiguities makes it hard for a system to reason about the query and to perform query evaluation.




Recent work in machine learning (ML), most notably Large Language Models (LLMs)~\cite{brown2020language} and Vision Language Models (VLMs)~\cite{bordes2024vlmsurvey}, has motivated a new class of \emph{multimodal} database management systems~\cite{patel2024lotus,chen2023symphony,yuan2024nsdb,urban2024caesura,urban2024eleet,shankar2024docetl, kurt2024xmode, xu2022eva, google2025bigquery, liu2025palimpzest}. Some of those systems ~\cite{xu2022eva, google2025bigquery} require users to manually compose SQL queries with optional ML user-defined functions (UDFs), providing flexibility yet remaining difficult for non‑expert users and tedious for experts. Other systems~\cite{patel2024lotus,chen2023symphony,yuan2024nsdb,urban2024caesura,urban2024eleet,shankar2024docetl, kurt2024xmode, liu2025palimpzest} delegate semantic interpretation entirely to foundation model operators.
At query time, they invoke models on each record by prompting it to generate the target attribute (e.g., an \texttt{is\_exciting} flag) 
and treat the model outputs as the final query result.
While straightforward, this paradigm has one major drawback:
The user receives only a final answer without information on how each tuple was derived because the generation process bypasses the relational layer. 

Users thus face a trade‑off: AI-assisted SQL engines that demand user effort, or powerful but opaque multimodal systems.
To reconcile these worlds, we introduce \sys\footnote{\textbf{kath}arós means clear and clean in Greek.}, an explainable \emph{multimodal DBMS} powered by LLM-driven human-AI collaboration.
Specifically, \sys makes the following three key contributions:

\noindent \textbf{(1) Unified semantic layer based on the relational model.} 
Unlike previous multimodal systems~\cite{xu2022eva,urban2024caesura,jo2024thalamusdb, chen2023symphony, yuan2024nsdb, liu2025palimpzest, urban2024eleet} that expose raw data to users, \sys introduces a unified \emph{relational} semantic layer of \emph{views} over data, giving the data a systematic, relational representation. 
This design has several advantages:
First, it unifies heterogeneous modalities under a single relational abstraction, enabling systematic, cost-based evaluation of cross-modal user queries. 
Second, the layer combines modality-specific powerful ML operators with the semantic guarantees of a traditional DBMS. 
Finally, the relational representation gives \sys the ability to track fine-grained lineage, allowing every result tuple to be traced back to its exact source records, and enabling better explainability.


\noindent \textbf{(2) Function-as-operator (FAO) query planning and execution.}
Similar to other multimodal systems~\cite{urban2024caesura, kurt2024xmode, chen2023symphony}, \sys transforms an NL request into a workflow of smaller steps. 
Each step corresponds to a transformation (e.g., similarity search) over data, or the combination of intermediate results from previous steps (e.g., join over views). 
This decomposition improves the understanding and verification of user intent. Importantly, \sys does this decomposition in three steps: it first converts the user NL query into a \textit{query sketch}, which is a step-by-step decomposition of the query, with a natural language description of the intent of each step (e.g., "Check the \texttt{Objects} table associated with each poster image to determine if a movie poster contains objects associated with excitement (e.g., weapons, motorcycles, etc.)"). 
Note that the keyword list is also generated by the LLM. During execution, however, if users are not satisfied with specific logic (e.g., the generated keyword list), they may provide feedback to \sys, which automatically updates the logic accordingly.
It then converts the query sketch into a logical plan where each node is a function, with a signature and description (e.g., 
"\texttt{gen\_excitement\_score()}"). 
Third, it generates the body of each function, associating a version identifier with each implementation (e.g., call a specific embedding model to embed the extracted objects in a poster image, embed the concepts from the generated keyword list, compute their similarity, and finally aggregate these values into an 'excitement' score per movie). 
This design has several advantages. 
First, it enables the system to explore different mappings between query sketch, logical plan, and physical plan (e.g., one step in a query sketch can correspond to multiple logical functions and logical function can further be decomposed in multiple physical functions or vice versa). 
Second, separating signature declaration from body synthesis lets \sys explore different interpretations for the same sub-task (e.g., interpreting "exciting movies" as action movies, recent releases, or award-winning movies). 
Third, splitting a query into small functions and generating each function separately reduces common problems in long generation such as hallucination~\cite{yang2025hallucinate} and error propagation~\cite{peng2024minorerror}.
Fourth, each function is assigned an identifier and a version tag, enabling the system to record how each output tuple is derived through a sequence of data transformations (e.g., when a new column is produced by an aggregation such as a count). Each transformation is implemented as a function, and these functions are persisted locally on disk. This gives \sys support for fine-grained lineage tracking, allowing it to trace any tuple in the final query result back through the intermediate materialized tables and the specific transformations (or functions) that produced it. 



Finally, the FAO design enables both rich logical rewrites and cost-based physical optimization. Recent systems for unstructured and multimodal data~\cite{patel2024lotus,russo2025abacus,Wei2025MultiObjectiveAR,shankar2024docetl} treat each semantic operator as a black-box prompt and search over models, prompts, or cascades to trade off cost and accuracy. 
In \sys, each FAO \emph{signature} serves as a logical operator, while each concrete implementation (e.g., prompting technique, coding variants) serves as a physical operator.
This separation allows \sys to attach cost and accuracy statistics to individual FAO implementations and compare alternatives for the same sub-task under a unified cost model, optimizing query accuracy and token cost subject to constraints (e.g., minimal user effort when rating sampled query results during plan profiling).
Because FAOs are composed into an explicit plan, the optimizer can jointly explore logical rewrites (e.g., predicate pushdown, fusing operators that share VLM calls) and physical choices (e.g., model cascades), much like traditional DBMS optimizers, but now over multimodal, model-driven operators.

\noindent \textbf{(3) Rich user interactions for query clarification, debugging, and explanation.} In \sys, we explore novel human-AI interactions for multimodal data management. 
Unlike traditional approaches to data management, user-system interaction does not have to be limited to a query-result pair: it can be iterative. 
Toward this goal, we build several dedicated channels for multi-turn clarifications between the user and \sys during the query interpretation, execution, and result explanation stages.
This design brings several advantages:
First, in the query parsing phase, vague user intent may cause \sys's interpretation to differ from the user's expectations. An interactive channel helps \sys better understand the query and draft more accurate plans.
Second, in the query execution phase,  unlike traditional DBMSs that abort on runtime errors, \sys fixes errors on-the-fly by exploring alternative function implementations and optionally involving users in debugging. 
Finally, after query execution, \sys enhances  transparency by explaining how a tuple or any intermediate result was derived using the lineage information described above. 
Thus, we envision that \sys will enhance query accuracy and strengthen user trust, despite using black-box LLMs as modules.

In summary, this vision paper makes the following contributions:
\begin{enumerate}
    \item We develop \sys, the first multimodal DBMS that combines traditional relational guarantees with modern AI operators and rich human-AI interactions to deliver trustworthy and explainable multimodal data management.
    \item We unify text, images, and video under a relational layer of views with version information, taking advantage of relational semantics while enabling fine-grained lineage across modalities.
    \item We introduce function-as-operator (FAO) query planning and execution, compiling each operator into a reusable and explainable function for modular and semantically flexible, multimodal pipelines.
    \item We provide conversational channels that let users interactively clarify NL queries, debug FAO implementations, and explain query results despite \sys containing LLM-powered modules.
\end{enumerate}

\sys takes an important step toward integrating AI models with DBMSs while ensuring explainability. This paper presents our vision, design, and preliminary results for \sys, which we are developing at the University of Washington. 



\section{\sys Architecture}
\label{sec:sys}
\begin{figure*}
    \centering
    \includegraphics[width=0.95\linewidth]{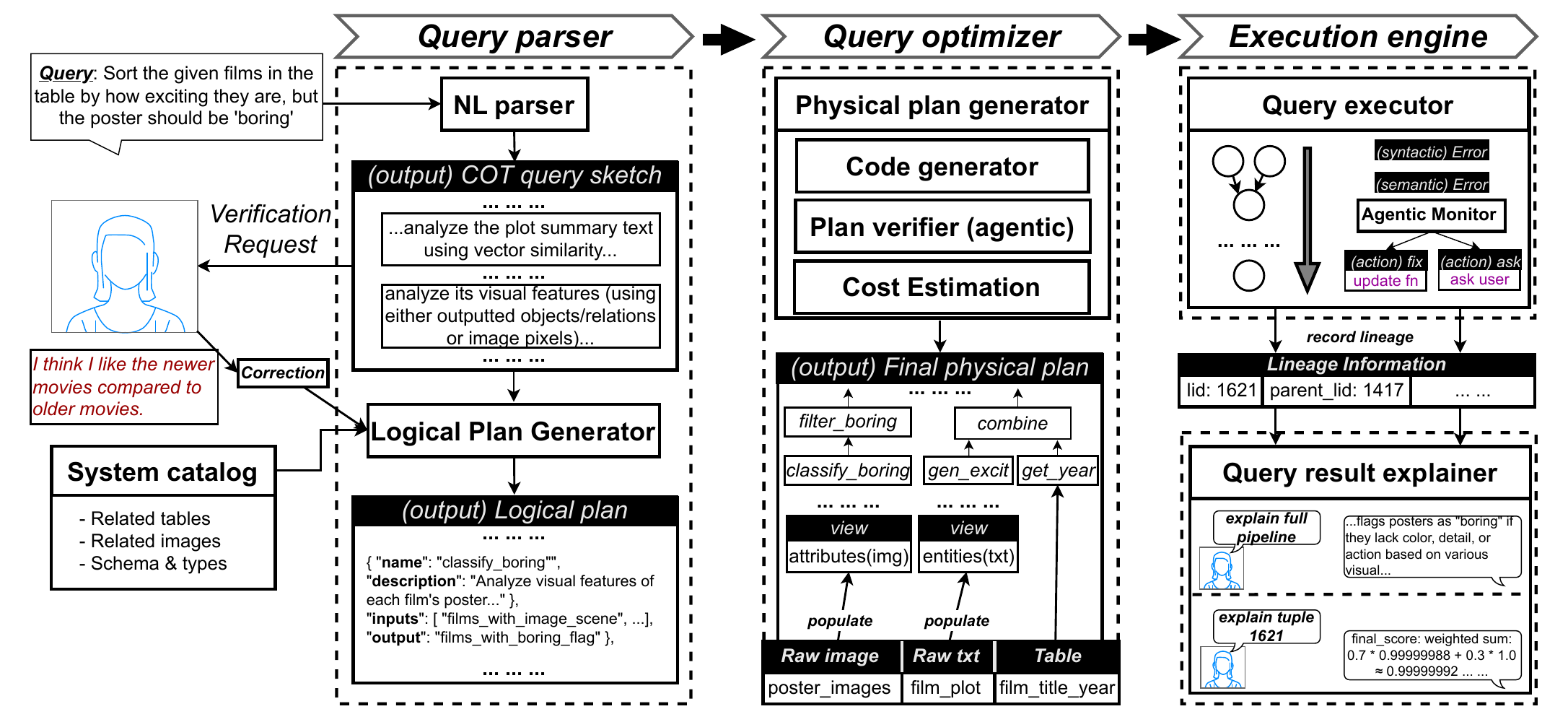}
    \caption{Overview of \sys.
    \sys consists of three main components: a query parser, an optimizer, and an execution engine.
    \sys accepts an NL query from the user as input. It generates a query sketch, then a logical plan, and then a physical plan, interacting with the user to seek clarifications as needed.
    \sys executes the physical query plan while fixing errors on-the-fly during execution and recording lineage information.
    Finally, \sys can explain query results at different granularities.}
    \label{fig:workflow}
\end{figure*}

We briefly describe the main components of \sys (Figure~\ref{fig:workflow}).


\subsection{Query Parser with Human-AI Verification}
A query parser in \sys\ converts a user’s NL request into an executable logical plan. There are two sub-modules within the query parser: \textbf{(1) NL parser.} Inspired by recent work showing that chain-of-thought (CoT) reasoning improves LLM performance~\cite{wei2022cot}, the \textit{NL parser} first generates a \textit{query sketch}, a step-by-step description of the intended execution logic expressed entirely in NL. The \textit{query sketch} deliberately avoids exposing operator-level details (e.g., function signatures or intermediate schemas) and thus remains one abstraction level above the final logical plan.
This higher-level representation is easier for users to inspect and edit as we discuss further in Section~\ref{sec:interactions}, yet still provides sufficient structure for the downstream compiler. \textbf{(2) Logical Plan Generator.} 
Given a query sketch as input, the \emph{logical plan generator} uses the system catalog as additional context and expands each step in the query sketch into a logical plan node equipped with a function signature. For example, given the query in Figure~\ref{fig:workflow}, and a query sketch step: \textit{"Analyze poster visual features using both extracted objects and image pixels to determine if the poster appears `boring' (e.g., lacks vivid colors, few objects, little action, plain background)"}, the logical plan generator produces a node in the logical plan shown in Figure~\ref{fig:lp-node} named \texttt{classify\_boring}.
We further describe the node's schema and its generation in Section~\ref{sec:fao}.


In the current prototype implementation, we assume a simple database schema containing the relevant tables and columns to use in the query. We are exploring extensions to more complex schemas, where \sys will need to automatically combine table lookups with similarity-based search to determine the relevant tables and columns to use in a query. We do not discuss this extension in this paper.




\subsection{Query Optimizer}

\sys's query optimizer translates a logical plan, in which each node contains only function signatures and schema-related information, into a low-cost physical plan, where the body of each function has been generated. 
A function can contain a SQL query over a table, a view population using machine learning models, a vector-based similarity search for semantic keyword matching, and more.
Since the \sys optimizer generates each function independently instead of producing the entire physical query plan at once, it reduces autoregressive error propagation~\cite{peng2024minorerror}. 
Additionally, it can generate these functions efficiently, in parallel. 
We describe the details of function generation in Section~\ref{sec:fao} and briefly discuss cost-based optimization with our proposed FAO paradigm.






\subsection{Execution Engine And Explainer}
\label{subsec:execution}

As shown on the right of Figure~\ref{fig:workflow}, \sys’s execution engine instantiates the physical plan, produces the result set, and exposes a channel for result explanations. 
Runtime errors are either \textit{syntactic}, which raises exceptions, or \textit{semantic}, where the LLM doubts that the output matches the user intent; \sys self‑repairs the former and seeks user clarification for the latter (Section~\ref{sec:interactions}). 
After execution, users can ask NL questions (e.g., how a particular tuple was derived or why an operator behaved as it did) about any intermediate tuple or the entire pipeline, a key capability enabled by our provenance model (Section~\ref{sec:dm}) and the interactive debugger (Section~\ref{sec:interactions}).

\section{\sys Data Model}
\label{sec:dm}

\begin{table}[t]
  \centering

  \begin{minipage}[t]{0.48\textwidth}
    \centering
    \small 
    \captionof{table}{Relational representation of image/video content. 
    }
    \label{tab:video-schema}
    \begin{tabular}{l}
      \toprule
      \textbf{Objects}(\underline{vid}, \underline{fid}, \underline{oid}, \textbf{lid}, cid, $x_1$, $y_1$, $x_2$, $y_2$) \\ \midrule
      \textbf{Relationships}(\underline{vid}, \underline{fid}, \underline{rid}, \textbf{lid}, oid$_i$, pid, oid$_j$) \\ \midrule
      \textbf{Attributes}(\underline{vid}, \underline{fid}, \underline{oid}, \textbf{lid}, k, v) \\
      \midrule
      \textbf{Frames}(\underline{vid}, \underline{fid}, \textbf{lid}, pixels) \\
      \bottomrule
    \end{tabular}
  \end{minipage}\hfill
   \newline

   \begin{minipage}[t]{0.48\textwidth}
    \centering
    \small 
    \captionof{table}{Relational representation of text content.}
    \label{tab:text-schema}
    \begin{tabular}{l}
      \toprule
      \textbf{Entities}(\underline{did}, \underline{eid}, \textbf{lid}, cid) \\ \midrule
      \textbf{Mentions}(\underline{did}, \underline{sid}, \underline{mid}, \textbf{lid}, eid, span$_1$, span$_2$) \\ \midrule
      \textbf{Relationships}(\underline{did}, \underline{sid}, \underline{rid}, \textbf{lid}, eid$_i$, pid, eid$_j$) \\ \midrule
      \textbf{Attributes}(\underline{did}, \underline{sid}, \underline{eid}, \textbf{lid}, k, v) \\
      \midrule
      \textbf{Texts}(\underline{did}, \textbf{lid}, chars) \\
      \bottomrule
    \end{tabular}
  \end{minipage}\hfill
  \newline
  
  \begin{minipage}[t]{0.48\textwidth}
    \centering
    \small 
    \captionof{table}{Unified provenance schema.}
    \label{tab:lineage-schema}
    \begin{tabular}{l}
      \toprule
      \textbf{Lineage}(\underline{lid}, \underline{parent\_lid}, src\_uri, func\_id, ver\_id, data\_type, ts)\\
      \bottomrule
    \end{tabular}
  \end{minipage}
\end{table}

In this section, we describe the data model \sys uses to align disparate data modalities under a unified relational schema. 
It has three main advantages: (1) it supports a variety of queries; (2) it provides a well-structured semantic foundation for queries; (3) it facilitates explainability and verification. Designing such a schema at the right level of granularity is non-trivial, as each modality has distinct characteristics (e.g., video combines spatial and temporal signals, whereas text may include multiple co-referring mentions to the same entity). An overly fine-grained schema may lead to high complexity, make view population expensive, and introduce attributes that may rarely matter during actual query execution. Conversely, an over-simplified schema may fail to capture essential semantics and thus might not reliably answer queries correctly. Requiring a user-defined schema adds unwanted burden, as our goal is for users to reason at the level of natural language. 
Our design must therefore achieve a balance: expressive enough to model modality-specific nuances, yet compact, tractable, and extensible to future modalities.

\noindent \textbf{Images and Videos as Scene Graphs.} 
Inspired by EQUI-VOCAL~\cite{zhang2023equi}, \sys adopts a scene graph~\cite{krishna2017visual} data model that represents visual content as objects interacting in space and time. Images are treated as videos with a single frame. Table~\ref{tab:video-schema} summarizes the relational schema that supports this model.
Video frames are uniquely identified by the pair \texttt{(vid, fid)}, which identify the video and the frame. An \textbf{Object} is defined as
\[
\texttt{(vid, fid, oid, lid, cid, x\_1, y\_1, x\_2, y\_2)},
\]
where \texttt{oid} is the unique id for the object, \texttt{cid} the class label (e.g., person), and \texttt{(x\_1, y\_1, x\_2, y\_2)} the upper-left and bottom-right bounding box coordinates. \texttt{lid} refers to the lineage, which we explain later in this section. An object can have \textbf{Relationships} with another object, defined as
\[
\texttt{(vid, fid, rid, lid, oid\_i, pid, oid\_j)}.
\]
Here, \texttt{rid} is a unique identifier for a relationship in the frame, and \texttt{pid} is the relationship class id between two objects.

Each object can also have \textbf{Attributes}
\[
\texttt{(vid, fid, oid, lid, k, v)},
\]
where \texttt{k} is the attribute key (e.g., \texttt{"color"}) and \texttt{v} is the attribute value (e.g., \texttt{"black"}).
Finally, the \textbf{Frames} table provides a view over the pixels within frames or images.
This simple yet powerful scene graph representation enables complex visual reasoning, even when the NL queries are indirect. For example, to find exciting movies, \sys may label two scenes as dangerous: one showing \emph{"a man jumped off a plane"} and the other \emph{"a dog fell into a pool"}. The scene graph allows \sys to explain why the latter does not make a movie “exciting.”

\noindent \textbf{Text content as text semantic graph.} 
Unlike standalone images, where each object is unique (or videos where each unique object can be tracked across frames), a textual corpus presents additional challenges, such as entity resolution~\cite{shankar2024docetl, christophides2020overview}. 
To address this challenge, we are experimenting with a schema that tries to capture the entity discussed in a document and the mentions that relate to those entities. 
Table~\ref{tab:text-schema} summarizes this text data model.
An entity from the \textbf{Entities} table corresponds to the system's best identification of individual entities in each document. 
Each entity can be mentioned multiple times and in different ways: for example through a full name (\texttt{''Taylor Swift''}), a pronoun (\texttt{''she''}), or an indirect reference (\texttt{''the artist behind the Eras tour''}). 
The schema for an entity is:
\[
\texttt{(did, eid, lid, cid)},
\]
where \texttt{did} is the document id, \texttt{eid} is the entity id shared by all mentions of that entity within the document, and \texttt{cid} is the entity class type (e.g., person).
\sys ensures \texttt{eid} uniqueness within the text corpus, but does not guarantee consistency across documents.
A mention from the \textbf{Mentions} table consists of a unique mention identifier (\texttt{mid}) occurring in a sentence (\texttt{sid}) of a given document (\texttt{did}), along with its character span (\texttt{span1, span2}). The schema is:
\[
\texttt{(did, sid, mid, lid, eid, span1, span2)},
\]
where \texttt{span1} and \texttt{span2} mark the start and end character positions of the mention.
For example, ``Taylor'' and ``Mrs.\ Swift'' will have two different \texttt{mid}'s but refer to the same entity ``Taylor Swift.'' (thus same \texttt{eid}). 
If a query treats mentions as independent objects without resolving them to their entities, it may yield incorrect results; grouping mentions by entity avoids such errors. 
Similar to the image scene-graph representation, \sys defines text \textbf{Relationships} (e.g., an entity \texttt{"Irwin Winkler"} can have relationship \texttt{"director\_of"} with another movie entity \texttt{"Guilty by Suspicion"}) and \textbf{Attributes} in key (e.g., \texttt{movie\_budget}) value (e.g., \texttt{13M}) format.
Their schemas follow the same pattern:
\[
\texttt{(did, rid, lid, eid\_i, pid, eid\_j)} \quad\text{for relationships,}
\]
\[
\texttt{(did, eid, lid, k, v)} \quad\text{for attributes.}
\]
Finally, the \textbf{Texts} table provides access to the raw textual content. The \texttt{lid} in every row is a unique lineage id which we discuss later.
It is worth noting that a text semantic graph is not the only way to represent unstructured text.  
In scientific papers, entire sentences expressing claims (e.g., ``our model outperforms prior work'') may serve as meaningful units, while legal documents often require finer granularity to track specific parties and events. 
\sys's relational schema is flexible to these variations, and an important research questions is to explore alternative designs and their impact on query accuracy.

\noindent \textbf{Provenance model.} \sys uses provenance to track how final and intermediate tuples are derived through a sequence of data transformations. 
Provenance is important in multimodal settings where derived information may come from heterogeneous sources (e.g., a bounding box from an image, a named entity from text, or a computed score). 
Users often need to understand \emph{why} a tuple appears in the result, and provenance enables \sys to provide grounded explanations of the derivation process.
As shown in Table~\ref{tab:lineage-schema}, each row records one edge in the provenance graph: a derived tuple (i.e., child) identifier (\texttt{lid}), an optional input tuple’s identifier (\texttt{parent\_lid}; NULL for external input data), an optional data path (\texttt{src\_uri}; e.g., an image object from an s3 bucket; NULL for intermediate tuples), a function identifier (\texttt{func\_id}) with a version number (\texttt{ver\_id}; as described in Section~\ref{sec:fao}) that produced the child at a certain timestamp (\texttt{created\_ts}), and a lineage data type (\texttt{data\_type}). 
Ingesting a raw table creates a single lineage entry with \texttt{data\_type}=\textsf{table}. 
For each function (including non-relational algebra ones), \sys also asks the same LLM that generates the function (Section~\ref{sec:fao}) to classify its \textit{dependency\_pattern} as one of four types: \textsf{one\_to\_one}, \textsf{one\_to\_many}, \textsf{many\_to\_one}, or \textsf{many\_to\_many}. 
The first two indicate single-tuple dependency, allowing \sys to record row-level lineage. 
In this case, the executor processes one input tuple at a time: when it reads an input tuple with \texttt{lid} and the function produces one or more output tuples, \sys sets each output tuple's \texttt{parent\_lid} to that input tuple's \texttt{lid}, assigns the output tuple a fresh \texttt{lid}, writes these fields into the tuple, and inserts a provenance entry reflecting this relationship with \texttt{data\_type}=\textsf{row}. 
The latter two indicate wide dependency (e.g., aggregation, sorting), for which \sys records only table-level lineage with \texttt{data\_type}=\textsf{table}, and assume that all input tuples have contributed to all the output tuples.
At present, \sys does not attempt to recover finer-grained multi-tuple provenance for wide-dependency operators.
\begin{figure}[t]
  \centering
  \includegraphics[width=0.85\linewidth]{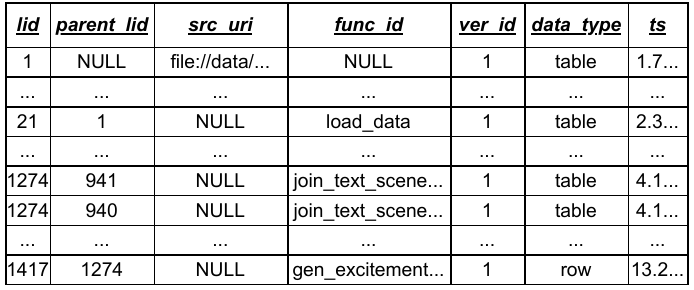}
  \caption{Example rows of a lineage table for an output tuple. 
  }
  \label{fig:pg}
\end{figure}
Figure~\ref{fig:pg} shows some sample rows from the lineage table for the query in Figure~\ref{fig:workflow}. 
One example lineage entry is the tuple \texttt{(lid=1417)}, whose \texttt{data\_type} is \texttt{row}. 
This tuple stores the excitement score for a movie, computed by the one-to-one function \texttt{gen\_excitement\_score}, which produces exactly one output row for each input row. 
Its parent is an intermediate result with \texttt{(lid=1274)}, whose \texttt{data\_type} is \texttt{table} and it is produced by the many-to-many operator \texttt{join\_text\_scene\_graph}. 
Here, the function \texttt{join\_text\_scene\_graph} joins two tables to associate each movie with the entities extracted from its plot description, and therefore its output is treated as a table-level artifact in the lineage graph. 
Accordingly, the tuple \texttt{(lid=1274)} has two parent tables, \texttt{(lid=940)} and \texttt{(lid=941)}, both of which were previously loaded into the system by other functions (omitted here for brevity).


Some important research questions related to \sys's provenance model are that lineage tracking adds a significant overhead, so how should \sys perform tracking without sacrificing much query execution speed? How fine-grained do we need the provenance to be in a multimodal DBMS?
Can \sys's multimodal schema together with lineage provide grounded explanations for complex query evaluation results based on user studies?

\section{Function-as-operator (FAO)}
\label{sec:fao}
 


After receiving a \textit{query sketch}, the \textit{logical plan generator} produces a \emph{logical plan} whose nodes are function signatures. 
The \textit{query optimizer} subsequently instantiates each signature with an executable function body (e.g., an SQL sub-query, a model-based inference routine that populates a view in the relational schema of Section~\ref{sec:dm}, and so on). 
Each function is stamped with a monotonically increasing \texttt{ver\_id}. 
Whenever the optimizer generates a new implementation for a function, \sys increments the version ID, leaving earlier versions intact. During execution, every output tuple carries the \texttt{ver\_id} of the function that produced it, enabling precise lineage queries, safe roll-backs to a prior version, and the iterative refinement workflows described in Section~\ref{sec:interactions}.
We refer to this design principle in \sys as \textbf{function-as-operator (FAO)}. There are a few challenges when adopting FAO in \sys.

\begin{figure}[t]
    \centering
    \includegraphics[width=0.99\linewidth]{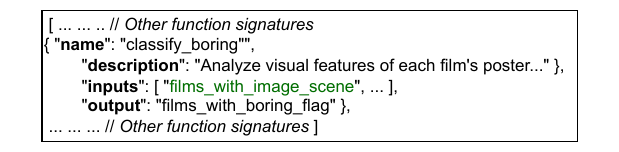}
    \caption{Function signature generated by the logical plan generator.}
    \label{fig:lp-node}
    \vspace{-0.1cm}
\end{figure}

\noindent\textbf{Generating function definitions with the \emph{logical plan generator}.}
To generate functions to evaluate the query, the logical plan generator first produces a tree of function signatures as the ``logical plan'', an example of which is shown in Figure~\ref{fig:lp-node}.
Each generated signature must contain the necessary information for the query optimizer to later generate the code while not being excessive to confuse the model~\cite{liuw2024lostinmiddle}.
Thus the logical plan generator produces each logical plan node strictly following our schema: every generated plan node is emitted in the \textit{exact JSON layout} we defined so the downstream parser can ingest it without any post-processing.
Here, \texttt{name} is the identifier of the function;
\texttt{inputs} is the list of datasource names that the function will consume (e.g., \texttt{classify\_boring} reads from a dataframe named "\texttt{films\_with\_image\_scene}").
A datasource may refer to (i) a base relation already materialized in \sys, whose schema is recorded in the catalog, or (ii) an intermediate table produced by a preceding node.
\texttt{output} declares the table produced by the function.
Finally, the \texttt{description} field provides semantic hints to support downstream code synthesis.

Inspired by the principles in~\cite{pan2025why}, we adopt a three-stage agentic workflow comprising a \emph{plan writer}, a \emph{tool user}, and a \emph{plan verifier}. 
The plan writer combines catalog metadata with the query sketch to draft a tree of logical-plan nodes. 
A verifier then reads the draft plan with the initial sample data (e.g., sample rows, column attributes, data types) from all related relations; 
if this snapshot is enough to judge correctness, it approves, otherwise it identifies \textit{specific relations} for which it needs additional information, invokes the tool user, which owns a small set of database utilities (e.g., rows sampler, joinability tester between two tables) to retrieve such information and judge again. 
Once the verifier is satisfied that the plan has realized the sketch, it forwards the logical plan to the query optimizer discussed next, otherwise it sends hints and the draft plan back to the writer to improve and review it again.

\noindent\textbf{Ensuring function executability with the \emph{query optimizer}.}
The optimizer implements each logical plan node.  
Nodes whose input does not depend on other nodes can be compiled in parallel, and for any given signature (e.g., Figure~\ref{fig:lp-node}) the optimizer may initialize multiple model instances to explore alternative implementations; our current prototype, however, implements functions sequentially.  
Three specialized agents collaborate on every node: a \emph{coder}, a \emph{profiler}, and a \emph{critic}.  
The optimizer first extracts column names, types, and sample rows from relations whose names are mentioned in a node's \texttt{inputs}.
Reading both the sampled rows and node specification, the coder writes a function body.
Then, the profiler uses the same set of sampled rows and executes the freshly generated function to ensure it can be executed and records its runtime for optimization purposes.  
Samples of intermediate results are provided to subsequent agents.  
If execution raises an exception, instead of aborting the entire query execution, \sys captures the stack trace, sampled data, parameters, and node metadata, and forwards them to the critic, which proposes a patch instead of aborting the query.

Unlike traditional relational systems, \sys must profile function implementations \emph{on-the-fly} during query execution, which can slow down the query. 
Although in practice this overhead is typically dominated by the LLM invocation time, an important research question is how \sys can reduce online profiling effort (e.g., through offline profiling) to speed up query plan generation. 



\noindent\textbf{Ensuring function semantic correctness with the \emph{query optimizer}.}
The same set of agents also checks that each function \textit{semantically} implements the logical node schema correctly. The critic first inspects the function source, sampled input records, produced output records, and node description to judge whether the results plausibly satisfy the intended semantics. For instance, in the workflow of Figure~\ref{fig:workflow}, a scoring function meant to generate a recency score based on user's request might be mistakenly implemented to do the reverse: giving higher score to the older movies.
When a mismatch is detected, the critic returns a corrective hint to the coder, which iterates until the output is acceptable. Extending this loop with optional human review for more complex edge cases is one of the research questions that we are exploring.

\noindent\textbf{Cost optimization with the \emph{query optimizer}.}
The optimizer can perform cost-based optimization at two levels. 
The first level is logical plan optimization. 
Here the optimizer may push predicates closer to data sources and merge two function signatures into one to avoid unnecessary intermediate result materialization. 
These rewrites shape the structure of the logical plan and influence how much work the system must perform later.
The second level is physical plan optimization. 
Here, each logical function signature can be implemented in multiple ways. 
For example, an image-to-text extraction operator may be instantiated using either a VLM-based implementation or an OCR-based implementation such as Tesseract\footnote{\url{https://github.com/tesseract-ocr/tesseract}}, each represented in \sys as a distinct function version (\texttt{ver\_id}). 
This versioning lets the optimizer choose among multiple concrete implementations of the same logical operator.
The optimizer profiles these implementations on sample input records and chooses the one that produces acceptable outputs at the lowest cost. 


Here, a critical challenge is that the query optimizer must balance cost against two goals: query accuracy and explainability.
A compact logical and physical plan with fewer larger functions may executes more quickly, but larger functions are more difficult to generate accurately, so such a plan may reduce accuracy. A smaller number of larger functions may also make explanations harder because little information about intermediate results is available. 
A more detailed plan can improve accuracy and provide clearer explanations, but it typically slows down execution. Exploring these trade-offs 
is an important research question. Additionally, as we discuss next, a user may provide comments and ask questions about query plans, which also opens the possibility of re-generating functions at different granularity if a user asks questions and needs to see additional intermediate results.

\section{Interactions}
\label{sec:interactions}

%



\begin{figure}[t]
    \centering
    \includegraphics[width=0.7\linewidth]{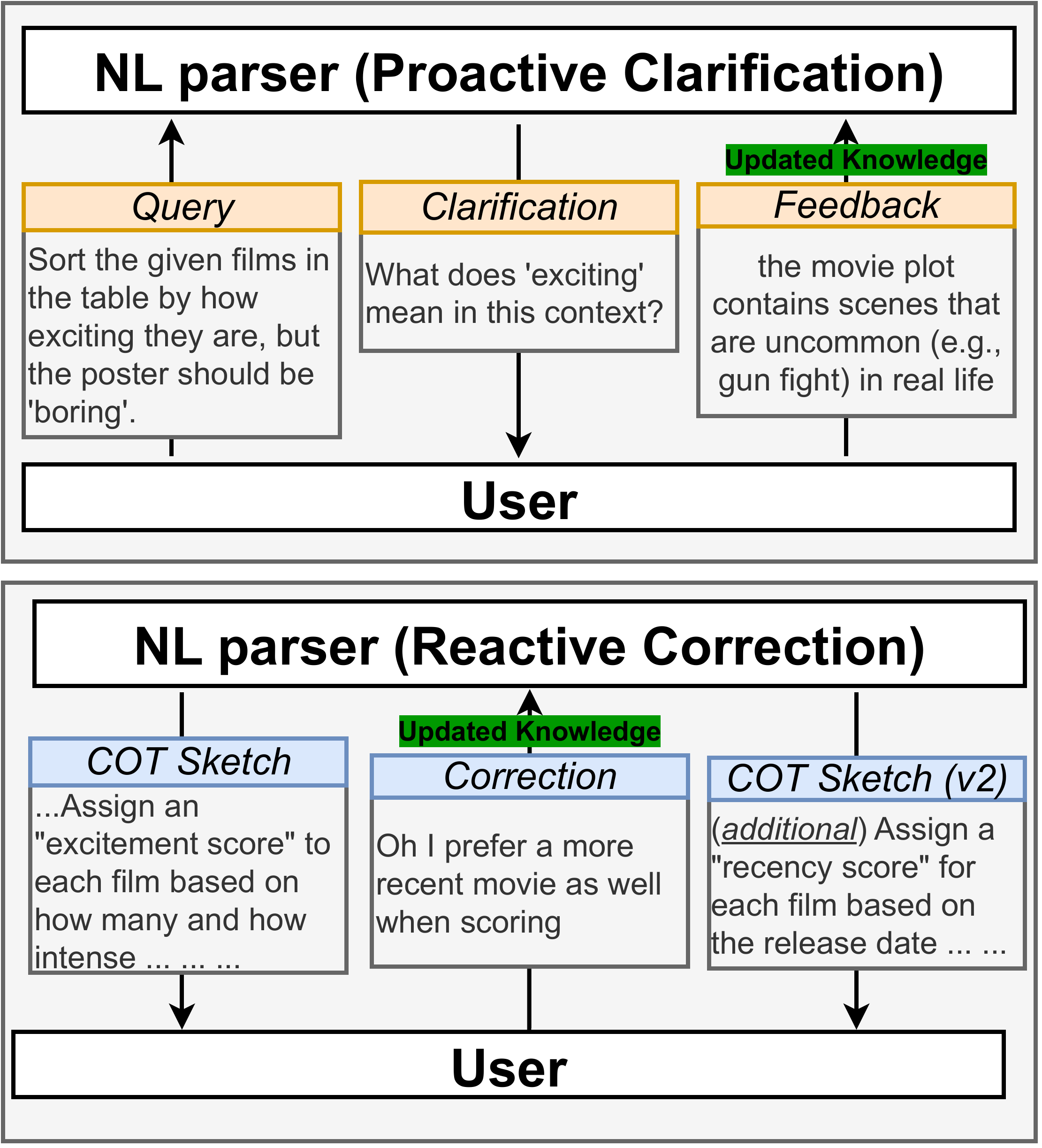}
    \caption{Example of NL parser interactions in two modes.}
    \label{fig:nlparser-ex}
    \vspace{-0.1cm}
\end{figure}

\noindent \textbf{Interactive NL Parser.} 
When the NL parser translates the user's potentially ambiguous NL query into a \textit{query sketch}, clarification or correction may be necessary. 
A straightforward but inefficient approach would be to show the user the entire query sketch and ask them to revise their original natural language request.
This approach, however, introduces unnecessary human effort as the entire query needs to be rewritten by the user.
Instead, in \sys, we experiment with two finer grained interaction models, and implement two collaborative agents: a \textit{reviewer} and a \textit{sketch generator}.




Inspired by recent work that advocates user involvement in the AI disambiguation process \cite{weld2019haii}, \sys's NL parser proactively asks clarification questions when it cannot confidently map a NL query to a single interpretation. As shown on the left of Figure~\ref{fig:workflow}, the reviewer agent first inspects the NL query and decides between two actions: (i) ask a clarification question if it detects unresolved ambiguity, or (ii) forward the request directly to the \textit{sketch generator}. Here, ambiguity depends on whether a term's meaning is context dependent or user dependent. 
For example, in Figure~\ref{fig:nlparser-ex} the word \textit{"exciting"} could be user-specific.
When such ambiguity is detected by the reviewer agent (prompted by \textit{"Look for ambiguous terms or subjective words...}"), it asks the user a focused question (e.g., \textit{"What does `exciting' mean in this context?"}).
The user then provides additional context, and the agent reassesses the query based on this newly provided information.
After generating the query sketch, the NL-parser performs reactive query correction based on user feedback. For example, the user may review the query sketch and realize that an important factor is missing (e.g., movie release year) (Figure~\ref{fig:nlparser-ex}, bottom, \texttt{Correction} box).
The user can tell the query writer to refine the sketch accordingly. The query writer incorporates the feedback, produces a revised sketch, and submits it for another round of review. 
This refinement cycle repeats until the user explicitly responds \texttt{OK}. 

An important research question is to comparatively evaluate these various feedback mechanisms and explore other possible ways of seeking user feedback at this stage of query execution to make the \textit{query sketch} as accurate as possible while minimizing user effort, as a \textit{query sketch} that does not match the user’s intent will inevitably lead to semantically incorrect functions being generated and erroneous final query results.

\noindent \textbf{Interactive Query Execution Debugger.} 
During execution, a function that cleared optimizer checks may still fail on unseen inputs.  
The monitor performs a role similar to the semantic checks at function-generation time, but now with the full input data. 

When the monitor detects a \textbf{syntactic fault} (e.g., unsupported file format), it launches a two-agent loop: the \textit{reviewer} diagnoses the exception, the \textit{rewriter} patches the code, increments its \texttt{ver\_id}, and resumes execution from the failed operator.  
Tuples unaffected by the error continue through the old function definition in a parallel process, preserving throughput. 
For example, the \texttt{classify\_boring} operator in Figure~\ref{fig:workflow} may rely on \texttt{cv2}\footnote{https://opencv.org/} to load and analyze image pixels to assess whether the poster's colors are vibrant, a feature that contributes to whether the image is 'boring' as specified in its function signature.
If it encounters an unsupported \textit{HEIC} file, the pipeline proceeds on other images while the rewriter adds a conversion step to a \texttt{cv2} compatible format.  

\textbf{Semantic anomalies} are subtler: the code runs but produces an outcome that the user does not expect. 
For example, a similarity-based vector join may mistakenly match the same poster image to several different movie titles, even though the code executes without error. 
In this case, the monitor inspects the resulting table, detects that a single poster image is linked to multiple movies, and flags this as unlikely to match the user’s intent. 
It then asks the user for confirmation or correction and, based on the feedback, updates the function logic so that the join behaves as intended. 
The monitor also explains a likely cause (e.g., the LLM may have implicitly assumed a one-to-one correspondence between poster images and tuples in \texttt{movie\_table}, an assumption that does not hold in practice~\cite{talmor2021multimodalqa} and produces spurious matches). 
To resolve the issue, the monitor prompts the user to either accept the operator as is, request an adjustment (e.g., enforce that each poster can be linked to only one tuple in \texttt{movie\_table}), or request a complete rewrite of the operator.

A research question at this stage of query execution is that an LLM-based \textit{monitor} examining intermediate results will incur additional token costs, so some type of sampling is necessary. Additionally, if a semantic error is found for the current tuple, then all the previous tuples must be reprocessed, and a question becomes whether any of the past work may be somehow leveraged.




\begin{figure}
    \centering
    \includegraphics[width=0.95\linewidth]{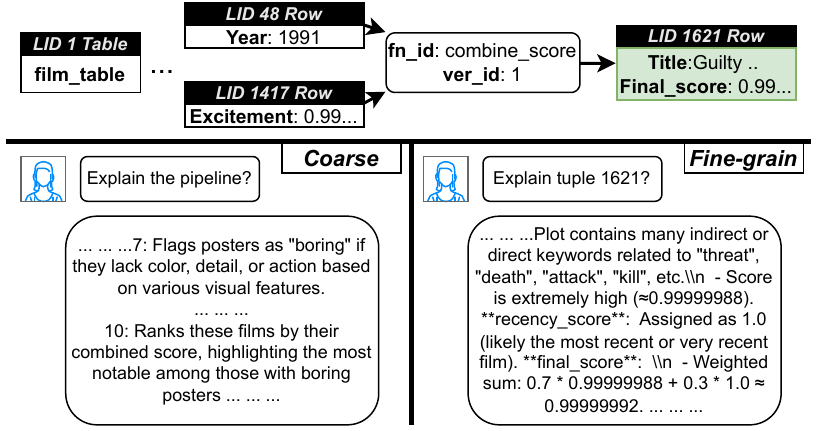}
    \caption{Example result of query explanations in two modes.}
    \label{fig:explanation}
    \vspace{-0.1cm}
\end{figure}

\noindent \textbf{Interactive Query Result Explanation.}
After the query execution stage, if a user receives any unexpected results, looking through the input data offers little insight and is prohibitively time-consuming on a large database.  
\sys overcomes this limitation by exposing the \emph{full provenance} of query results and \textit{makes it queryable in NL}: For every output tuple, \sys can show the materialized view it came from (described in Section~\ref{sec:dm}), how it was derived by the pipeline of FOAs, and how each function may have been updated during the agentic query-generation process. 
Additionally, the user can also ask NL queries over this lineage information as shown in Figure~\ref{fig:explanation}.
\sys supports two explanation modes. 
The coarse-grained mode shows a high‑level overview of the transformations performed during query execution (e.g., showing that \sys decides whether a poster is "boring" by analyzing its visual features and the objects depicted).
The fine‑grained mode offers a low-level explanation: it takes a specific \texttt{lid} as input, inspects the function signature and implementation, traces parent tuples, and shows the details of how every output tuple field was derived (e.g., showing that the final score combines a recency score of 1 with an excitement score from earlier functions, each traceable by the user). Overall, our intermediate relational view layer and fine-grained lineage enable \sys{} to provide more detailed and consistent explanations than larger-scoped, black-box LLM invocations.

An important research question here is that query execution over very large databases produces large intermediate results and large lineage information. 
How should the system best summarize this information and provide answers to users' query explanation questions efficiently and at low cost? Can the system not save all the lineage in the first place but only sample lineage information?

\section{Initial Result}
\label{sec:res}
\begin{figure}[t]
    \centering
    \includegraphics[width=0.9\linewidth]{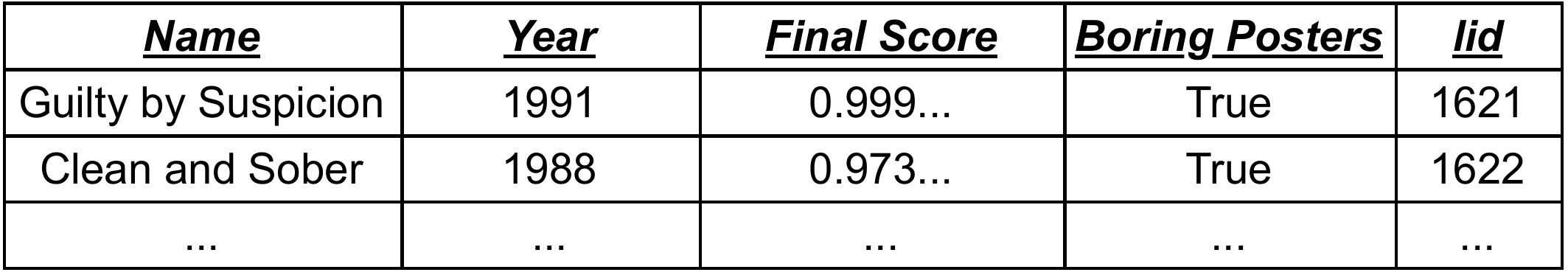}
    \caption{Example final output of \sys.}
    \label{fig:res}
\end{figure}

We execute the example query shown in Figure~\ref{fig:workflow} with \sys over MMQA~\cite{talmor2021multimodalqa}, a dataset that contains tables, texts, and images crawled from Wikipedia. 
The top two results are shown in Figure~\ref{fig:res}.
The query parser accepts the query and asks the following clarification question: \textit{"What does `exciting' mean in this context?"} 
We simulate the following user reply: \textit{"The movie plot contains scenes that are uncommon in real life" }(Figure~\ref{fig:nlparser-ex}); the parser then generates a query sketch with eight steps. 
We further simulate the user adding an additional requirement, \textit{"I prefer more recent movies when scoring,"}. The parser updates the plan and produces an $11$-step query sketch.
For the current prototype, we have pre-written the view-population function that invokes GPT-4o and supplies schema information to \sys as the first step, leaving $10$ remaining logical plan nodes.
The query plan optimizer generates the following functions:
(1) selects the relevant columns from \texttt{movie\_table} (e.g., title, release year); (2) joins the relational view over text with \texttt{movie\_table}; (3) joins the relational view over images with \texttt{movie\_table}; (4) computes excitement scores by measuring vector similarity between keywords (e.g., \emph{gun}, \emph{murder}, ...) and all extracted text entities (Note: it is worth noting that a LLM generates the keyword list here); (5) assigns recency scores based on the release year; (6) combines excitement and recency scores according to user request; (7) classifies each poster as boring or not using the raw image and its scene-graph view; (8) filters out posters labeled as boring; and (9) and (10) joins all intermediate results to produce the final ranked list of movies by their combined score.
Finally, a tuple (\texttt{lid=1621}) is generated, as shown in Figure~\ref{fig:res}. The user then requests result explanations for the entire pipeline (Figure~\ref{fig:explanation}, left) and for how the final tuple is produced (right). Only a snippet is shown due to space constraints.

\section{Related Work}
\label{sec:rw}


Recent work has focused on building data systems that support queries over multimodal data.
These systems can be broadly grouped into two categories: 
One set of systems~\cite{xu2022eva, jo2024thalamusdb, google2025bigquery, liu2025palimpzest, urban2024eleet, patel2024lotus} consider ML models (e.g., an object classifier) as operators and requires users to write SQL or Python code explicitly. 
Instead, \sys accepts NL queries, performs iterative and interactive query refinement, and leverages LLMs both as a query planner and as a function generator. 
A second line of work~\cite{urban2024caesura, kurt2024xmode, chen2023symphony} also takes NL queries and uses LLMs \textit{as black-boxes} to plan and execute them. 
However, \sys differs from these systems in two key ways: (i) it supports richer communication with the user, enabling iterative query parsing and execution, and (ii) it unifies diverse modalities behind a single, relational semantic layer, improving query semantics and explainability.
A third line of work~\cite{arora2023language, chen2023seed} proposes approaches that generate functions for populating relational views or performing data curation tasks over single-modality unstructured text. 
\sys differs from them in that it supports multimodal data, generates functions not only for view population but also for query execution, and includes a lineage tracking system for explainability.




\section{Conclusion}
\label{sec:concl}
We introduced \sys, a multimodal DBMS with explainable query execution. \sys combines multimodal data under a unified relational view, implements a new FAO model, and keeps users in the loop through interactive channels for clarification, correction, execution guidance, and result explanation.

\begin{acks}
This work was supported in part by the NSF through awards 2211133. We thank the reviewers for the helpful comments.
\end{acks}

\bibliographystyle{ACM-Reference-Format}
\bibliography{dbms, ai}

@article{krishna2017visual,
  title={Visual genome: Connecting language and vision using crowdsourced dense image annotations},
  author={Krishna et al.},
  journal={International journal of computer vision},
  volume={123},
  pages={32--73},
  year={2017},
  publisher={Springer}
}

@article{wei2022cot,
  title={Chain-of-thought prompting elicits reasoning in large language models},
  author={Wei et al.},
  journal={Advances in neural information processing systems},
  volume={35},
  pages={24824--24837},
  year={2022}
}

@article{peng2024minorerror,
  title={Stepwise reasoning error disruption attack of llms},
  author={Peng et al.},
  journal={arXiv preprint arXiv:2412.11934},
  year={2024}
}

@article{yang2025hallucinate,
  title={Hallucinate at the Last in Long Response Generation: A Case Study on Long Document Summarization},
  author={Yang et al.},
  journal={arXiv preprint arXiv:2505.15291},
  year={2025}
}

@article{liuw2024lostinmiddle,
    title = "Lost in the Middle: How Language Models Use Long Contexts",
    author = "Liu et al.",
    journal = "Transactions of the Association for Computational Linguistics",
    volume = "12",
    year = "2024",
    address = "Cambridge, MA",
    publisher = "MIT Press",
    url = "https://aclanthology.org/2024.tacl-1.9/",
    doi = "10.1162/tacl_a_00638",
    pages = "157--173",
}

@inproceedings{pan2025why,
title={Why Do Multiagent Systems Fail?},
author={Pan et al.},
booktitle={ICLR 2025 Workshop on Building Trust in Language Models and Applications},
year={2025},
url={https://openreview.net/forum?id=wM521FqPvI}
}

@article{bordes2024vlmsurvey,
  title={An introduction to vision-language modeling},
  author={Bordes et al.},
  journal={arXiv preprint arXiv:2405.17247},
  year={2024}
}

@article{brown2020language,
  title={Language models are few-shot learners},
  author={Brown et al.},
  journal={Advances in neural information processing systems},
  volume={33},
  pages={1877--1901},
  year={2020}
}

@inproceedings{talmor2021multimodalqa,
title={MultiModal{\{}QA{\}}: complex question answering over text, tables and images},
author={Talmor et al.},
booktitle={International Conference on Learning Representations},
year={2021},
url={https://openreview.net/forum?id=ee6W5UgQLa}
}

@inproceedings{weld2019haii,
author = {Amershi et al.},
title = {Guidelines for Human-AI Interaction},
year = {2019},
isbn = {9781450359702},
publisher = {Association for Computing Machinery},
address = {New York, NY, USA},
url = {https://doi.org/10.1145/3290605.3300233},
doi = {10.1145/3290605.3300233},
booktitle = {Proceedings of the 2019 CHI Conference on Human Factors in Computing Systems},
pages = {1–13},
numpages = {13},
location = {Glasgow, Scotland Uk},
series = {CHI '19}
}

@article{zhang2023equi,
  title={Equi-vocal: Synthesizing queries for compositional video events from limited user interactions},
  author={Zhang et al.},
  journal={Proceedings of the VLDB Endowment},
  volume={16},
  number={11},
  pages={2714--2727},
  year={2023},
  publisher={VLDB Endowment}
}

@article{yuan2024nsdb,
  title={nsdb: Architecting the next generation database by integrating neural and symbolic systems},
  author={Yuan et al.},
  journal={Proceedings of the VLDB Endowment},
  volume={17},
  number={11},
  pages={3283--3289},
  year={2024},
  publisher={VLDB Endowment}
}

@inproceedings{chen2023symphony,
  title={Symphony: Towards Natural Language Query Answering over Multi-modal Data Lakes.},
  author={Chen et al.},
  booktitle={CIDR},
  pages={1--7},
  year={2023}
}

@inproceedings{xu2022eva,
  title={EVA: A symbolic approach to accelerating exploratory video analytics with materialized views},
  author={Xu et al.},
  booktitle={Proceedings of the 2022 International Conference on Management of Data},
  pages={602--616},
  year={2022}
}

@inproceedings{urban2024caesura,
  title={Demonstrating CAESURA: Language Models as Multi-Modal Query Planners},
  author={Urban et al.},
  booktitle={Companion of the 2024 International Conference on Management of Data},
  pages={472--475},
  year={2024}
}

@article{kurt2024xmode,
  title={Explainable Multi-Modal Data Exploration in Natural Language via LLM Agent},
  author={Nooralahzadeh et al.},
  journal={arXiv preprint arXiv:2412.18428},
  year={2024}
}

@article{patel2024lotus,
  title={Semantic Operators: A Declarative Model for Rich, AI-based Data Processing},
  author={Patel et al.},
  journal={arXiv preprint arXiv:2407.11418},
  year={2024}
}

@inproceedings{chaudhuri1998overview,
  title={An overview of query optimization in relational systems},
  author={Chaudhuri, Surajit},
  booktitle={Proceedings of the seventeenth ACM SIGACT-SIGMOD-SIGART symposium on Principles of database systems},
  pages={34--43},
  year={1998}
}

@inproceedings{liu2025palimpzest,
  title={Palimpzest: Optimizing ai-powered analytics with declarative query processing},
  author={Liu et al.},
  booktitle={Proceedings of the Conference on Innovative Database Research (CIDR)},
  pages={2},
  year={2025}
}

@misc{google2025bigquery,
  author       = {{Google Cloud}},
  title        = {{BigQuery}: Fully Managed Data Warehouse},
  year         = {2025},
  howpublished = {\url{https://cloud.google.com/bigquery}},
  note         = {Accessed 23 Jul 2025}
}

@article{jo2024thalamusdb,
  title={Thalamusdb: Approximate query processing on multi-modal data},
  author={Jo et al.},
  journal={Proceedings of the ACM on Management of Data},
  volume={2},
  number={3},
  pages={1--26},
  year={2024},
  publisher={ACM New York, NY, USA}
}

@article{shankar2024docetl,
  title={DocETL: Agentic Query Rewriting and Evaluation for Complex Document Processing},
  author={Shankar et al.},
  journal={arXiv preprint arXiv:2410.12189},
  year={2024}
}

@article{christophides2020overview,
  title={An overview of end-to-end entity resolution for big data},
  author={Christophides et al.},
  journal={ACM Computing Surveys (CSUR)},
  volume={53},
  number={6},
  pages={1--42},
  year={2020},
  publisher={ACM New York, NY, USA}
}

@article{urban2024eleet,
  title={Eleet: Efficient learned query execution over text and tables},
  author={Urban et al.},
  journal={Proceedings of the VLDB Endowment},
  volume={17},
  number={13},
  pages={4867--4880},
  year={2024},
  publisher={VLDB Endowment}
}

@article{arora2023language,
  title={Language Models Enable Simple Systems for Generating Structured Views of Heterogeneous Data Lakes},
  author={Arora et al.},
  journal={Proceedings of the VLDB Endowment},
  volume={17},
  number={2},
  pages={92--105},
  year={2023},
  publisher={VLDB Endowment}
}

@article{chen2023seed,
  title={SEED: Domain-specific data curation with large language models},
  author={Chen et al.},
  journal={arXiv preprint arXiv:2310.00749},
  year={2023}
}

@article{russo2025abacus,
  title={Abacus: A Cost-Based Optimizer for Semantic Operator Systems},
  author={Russo et al.},
  journal={arXiv preprint arXiv:2505.14661},
  year={2025}
}

@inproceedings{Wei2025MultiObjectiveAR,title={Multi-Objective Agentic Rewrites for Unstructured Data Processing},
author={Lindsey Linxi Wei et al.},
year={2025},
url={https://api.semanticscholar.org/CorpusID:283458157}
}

\end{document}